\documentclass[atoms,article,submit,pdftex,moreauthors]{mdpi} 

\firstpage{1} 
\pubvolume{1}
\issuenum{1}
\articlenumber{0}
\pubyear{2024}
\copyrightyear{2024}
\datereceived{ } 
\daterevised{ } 
\dateaccepted{ } 
\datepublished{ } 

\hreflink{https://doi.org/} 

\usepackage{subfig}
\usepackage{graphicx}
\usepackage{float}
\usepackage{natbib}

\Title{Many-body effects in a composite bosonic Josephson junction}

\Author{Sudip Kumar Haldar $^{1,}$\orcidA{} and Anal Bhowmik $^{2,3,\dagger}$*\orcidB{}}

\address{%
$^{1}$ \quad Department of Physics and Material Science $\&$ Engineering, Jaypee Institute of Information Technology, Noida, India 201304; sudip.haldar@jiit.ac.in\\
$^{2}$ \quad Department of Physics, University of Haifa,  Haifa, Israel 3498838;\\
$^{3}$ \quad Haifa Research Center for Theoretical Physics and Astrophysics, University of Haifa,  Haifa, Israel 3498838; ab.physics2003@gmail.com}

\firstnote{Current address: Homer L. Dodge Department of Physics and Astronomy, The University of Oklahoma, Norman, Oklahoma 73019, USA.}

\corres{Correspondence: ab.physics2003@gmail.com}  

\abstract{In standard bosonic Josephson junctions (BJJs), particles tunnel between two single-well potentials linked by a finite barrier. The dynamics of standard BJJs have been extensively studied, both at the many-body and mean-field levels of theory. In the present work, we introduce the concept of a composite BJJ. In a composite BJJ, particles tunnel between two double-well potentials linked by a finite potential barrier between them. We focus on many-body facets of quantum dynamics and investigate how the complex structure of the junction influences tunneling. Employing the multiconfigurational time-dependent Hartree for bosons method, highly-accurate many-boson wave functions are obtained from which properties are computed. We analyze the dynamics using the survival probability, the degree of fragmentation of the junction, and the fluctuations of observables, and discuss how many-boson tunneling behaves, and how it may be controlled, using the composite nature of the junction. A central result of this work relates to the degree of fragmentation of composite BJJs with different numbers of bosons. We provide strong evidence that a universal degree of fragmentation into multiple time-dependent modes takes place. Further applications are briefly discussed.}

\keyword{Josephson junction; many-body theory; fragmentation} 

\begin{document}

\setcounter{section}{0} 

\section{Introduction}

Understanding the quantum tunneling of correlated systems is a fundamental problem of quantum mechanics since it is the main mechanism behind several phenomena observed in different disciplines of physics,  starting from condensed matter physics to quantum information and communication~\cite{Cormick2023, tao2023arxiv, benatti2023arxiv, lewis2023arxiv}. It is imperative to have a highly maneuverable system to explore the various features of many-body quantum tunneling and develop a deep understanding of the physics of many-body quantum tunneling of correlated systems.  In this context, it is to be noted that ultra-cold atomic systems and quantum gases have emerged as one of the preferred choices for the simulation of correlated quantum many-body systems. 

In particular, trapped ultra-cold bosons in a symmetric double-well potential called the bosonic Josephson junction (BJJ) \cite{Sakmann2009, Haldar2018, Bhowmik2020},  provide a prototype for the Josephson effect originally predicted for tunneling of  Cooper pair between two weakly-linked superconductors \cite{Josephson1962}. Therefore, BJJ provides a deep insight into the tunneling dynamics of correlated quantum many-body systems.  Naturally, it has been the focus of a large number of theoretical as well as experimental research~\cite{Smerzi1997, Albiez2005, Gati2007, LeBlanc2011, Gillet2014, Burchinati2017, Hou2018}. The mean-field theory and the two-mode Bose-Hubbard (BH) model are commonly used for theoretically studying BJJ. While several important features of BJJ dynamics, such as the collapse and revival of the density oscillations~\cite{Milburn1997} and the fragmentation~\cite{Sakmann2014, Theel2020, Vargas2021}, can not be captured by the mean-field level of theory, BH model can at best provide a qualitative description for such features and ultimately totally fails to capture the many-body features of the many-particle variances. This is due to not taking into account the role of the higher energy band. Therefore to accurately describe the BJJ dynamics, in particular, and the dynamics of a correlated many-body system in general, at least when the higher energy bands participate, solving the many-body Schr\"odinger equation is necessary. In this context, our group has developed a numerically exact many-body method, called the Multiconfigurational time-dependent Hartree method for bosons (MCTDHB),  which takes into account all the participating energy bands.

In recent times, MCTDHB has been thoroughly utilized in studying the BJJ dynamics in different situations, viz., in 1D as well as in 2D; with repulsive and attractive interactions \cite{Sakmann2010}; with contact and finite range interactions \cite{Tunneling_Rapha, Haldar2018, Bhowmik2022arxiv}; and in presence of an asymmetry in the trap \cite{Haldar2019, Bhowmik2022}, etc. Studying fragmentation and the uncertainty product of the many-particle position and momentum operators by MCTDHB~\cite{Klaiman2016} has only reinforced the importance of using a full many-body model. 

The many-particle uncertainty product of the BEC in a double well was shown to grow with time $t$ as $t^2$ (up to leading order in $t$). It has been further shown that, contrary to the BH dimer, the full many-body dynamics of BJJ for the repulsive and attractive interactions are not equivalent \cite{Sakmann2010}.  Also, the fragmentation of BEC in both the symmetric and asymmetric double-well trap has been shown to exhibit an indifference to the particle number $N$ as long as the interaction parameter $\Lambda =\lambda_0 (N-1)$ ($\lambda_0$ being the strength of interaction) remains unchanged. Moreover, resonantly enhanced tunneling, which was experimentally observed, has also been demonstrated for an asymmetric BJJ with MCTDHB \cite{Haldar2019}.  On a different note, the Josephson effects have been investigated in various complex systems,  such as two-component BEC \cite{Mondal2022}, spinor condensates \cite{Zibold2010}, polariton condensates \cite{Abbarchi2013}, fermionic superfluid~\cite{Valtolina2015}, and spin-orbit coupled BEC \cite{Hou2018}.

Although various intriguing many-body effects have already been observed in one-dimensional ultra-cold bosonic ensembles, in all those studies, only the lowest band had a significant role in the investigated parameter regime.  To explore the role of higher energy bands in the many-body dynamics in one spatial dimension, in this work,  we investigate the many-body dynamics of an interacting bosonic system in a composite double well which we call a composite bosonic Josephson junction (CBJJ). In CBJJ, the system tunnels between two double-well potentials connected via a barrier of finite height \textcolor{black}{which may be used to model Josephson heat oscillations~\cite{Strzys}}. The complex geometry of the system makes the many-body tunneling richer with the involvement of the higher energy bands. \textcolor{black}{Tunneling dynamics of BEC in multiwell potentials are also important due to its potential applications in atomtronics-enabled quantum technologies~\cite{amico_2017, amico_2021, RMP2022} as well as for providing a prototype for nonintegrable systems~\cite{Vittorio_2003, Doron_2018, Santos_2022}. The dynamics of BEC in a four-well \cite{four-well1, four-well2, KARMAKAR} trap has already been attempted theoretically. Trapping potential with double wells~\cite{Folling2007, Gati2007} and triple wells~\cite{Caliga_2016} as well as optical lattice~\cite{Bloch2005} (which may be conceived as a collection of shallow wells) are routinely realized in experiments. By the same token, it should be possible to experimentally realize composite double-well traps with the available technology. In particular, a quadrupole well system can be realized either by using four closely neighboring microtrap potentials or with a double-well trap in which two internal states are coupled by a Rabi laser~\cite{PRL1999, PRL1997}.} 

To highlight the extent of the participation of the higher energy bands, we analyzed the variation of survival probability, occupation numbers, and fluctuations as a function of time. Already for BJJ,  the survival probability is directly related to experimentally observed population imbalance and also theoretically studied by a mean-field method. However, experimentally observed features of decay in the oscillations of population imbalance cannot be described by the mean-field method and requires many-body theory such as MCTDHB~\cite{Haldar2019}. It is obvious that fragmentation and depletion defined through the occupation numbers in higher orbitals can only be studied by a many-body method. Similarly, it is not possible even to capture the qualitative features of the Many-particle position and momentum variances which are measures of fluctuations in the system~\cite{Klaiman2016}. Therefore, all the quantities considered here have already shown distinct many-body features for BJJ and are expected to show more prominent many-body signatures for CBJJ where the higher orbitals are expected to have a greater role due to the complexity of the geometry of the system. 

The organization of the paper is as follows. Section~\ref{method} introduces the system studied here and gives an outline of the methodology used. We discuss our findings in section~\ref{result}. We summarise our main findings and draw our conclusions in section~\ref{conclusion}.  Further details of the methodology as well as the discussion about the numerical convergence of our results are discussed in the Appendix.

\section{Formalism}~\label{method}
\begin{figure}[!h]
\subfloat[]{\includegraphics[width=0.30\textwidth]{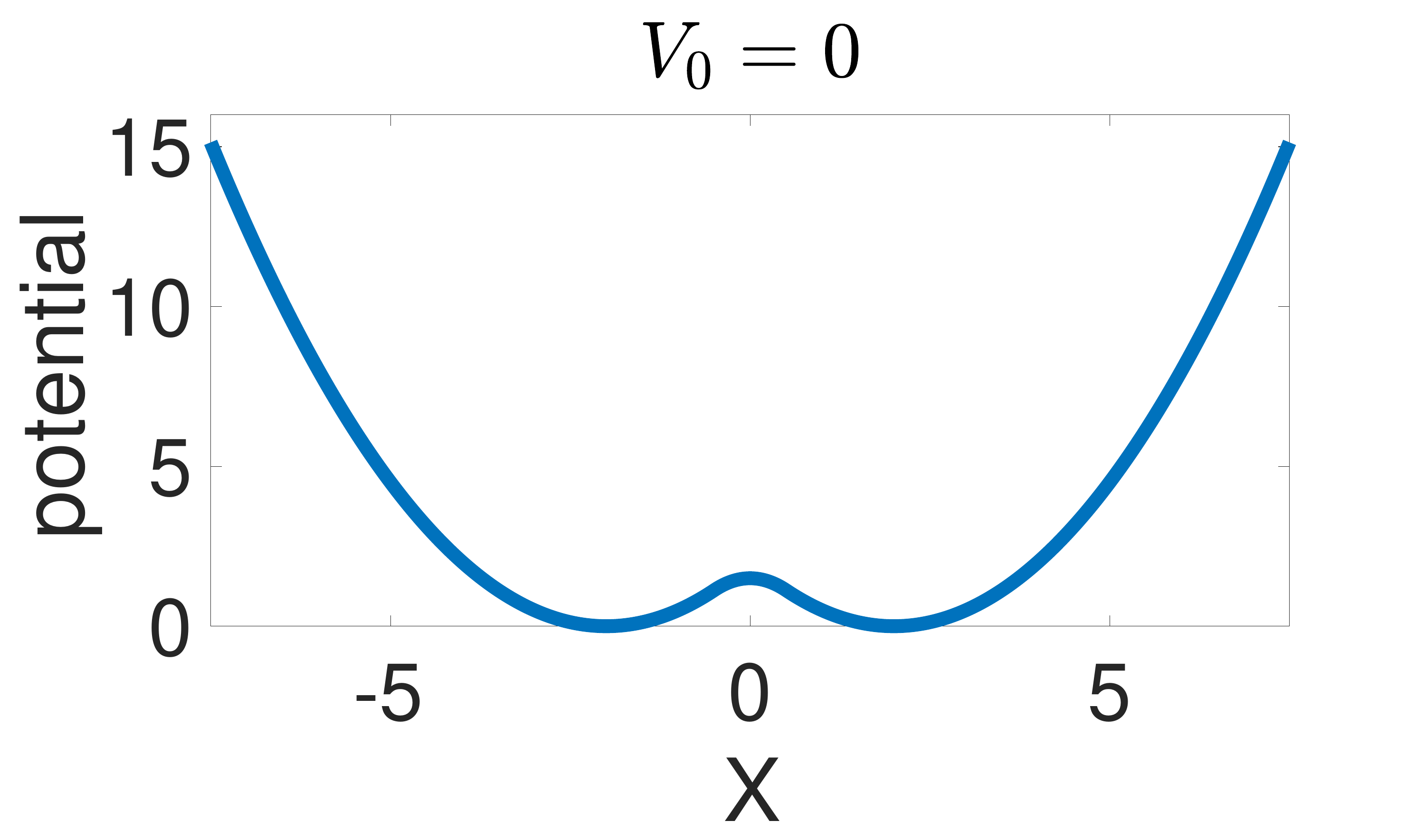}}
\subfloat[]{\includegraphics[width=0.30\textwidth]{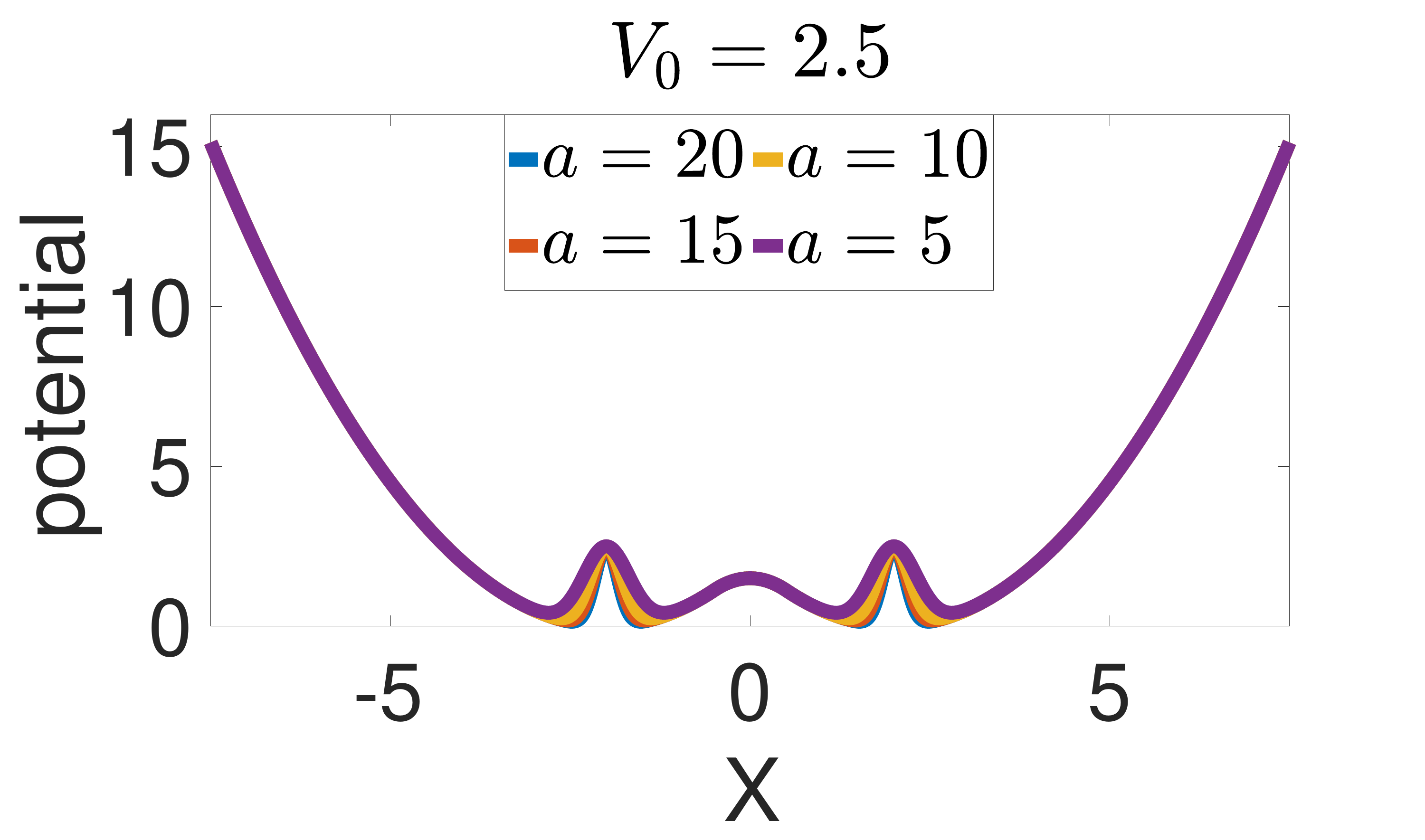}}  
 \subfloat[]{\includegraphics[width=0.30\textwidth]{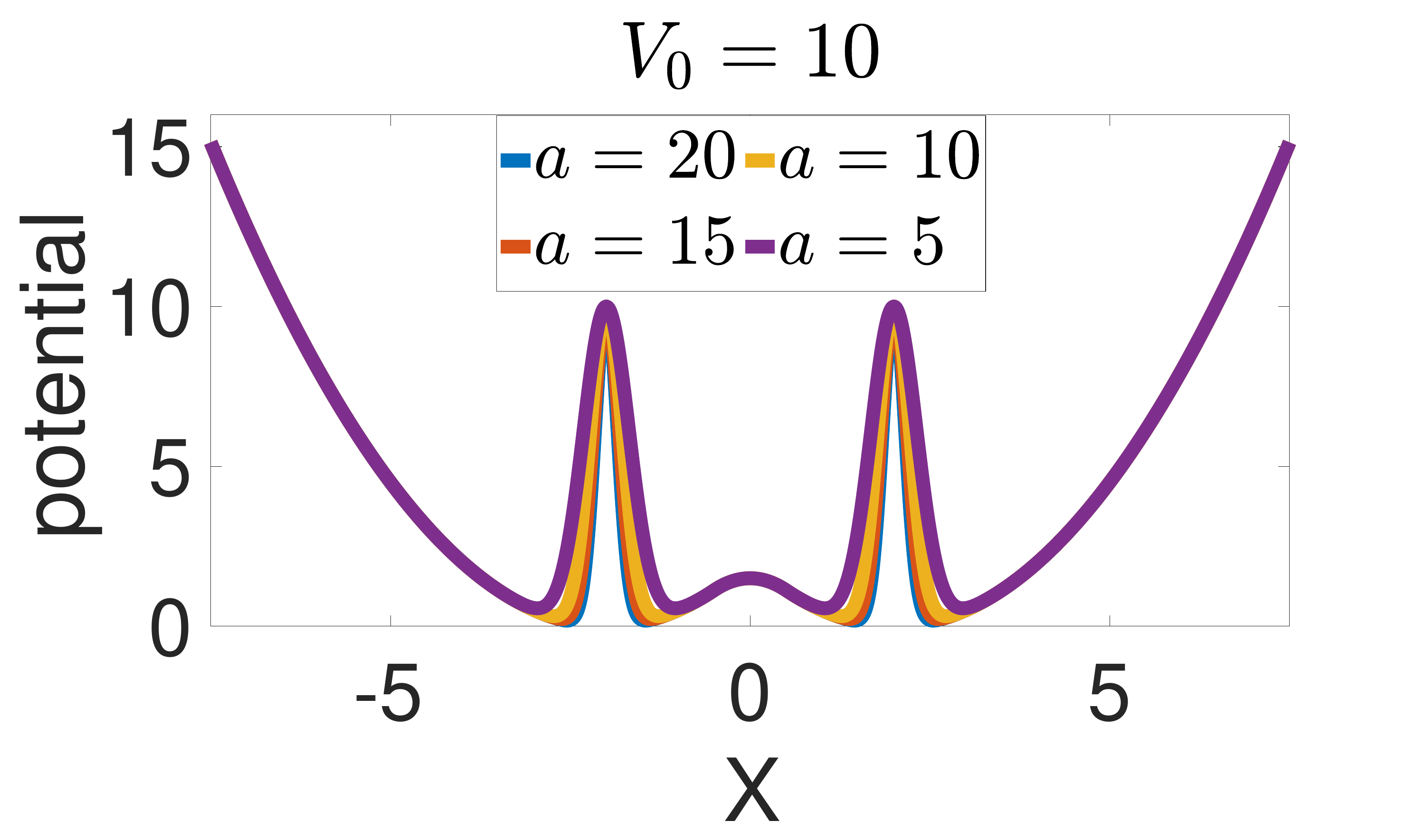}}   \\
\caption{External trapping potential presented in Eq.~\ref{eq-dyn} for (a) $V_0=0$, (b) $V_0=2.5$, and  (c) $V_0=10$. At $V_0=0$, the trapping potential is a regular double well and with ramping up the barrier height $V_0$, the trapping potential becomes a composite double-well. $a$ controls the width of the intra-well barrier as shown in (b) and (c) for $V_0=2.5$ and $V_0={10}$, respectively. The quantities shown here are dimensionless.}
\label{fig-trap}
\end{figure}

To explore the role of the higher energy bands in the many-body dynamics of correlated systems, here we considered the post-quench many-body tunneling dynamics of a system of interacting ultra-cold bosons trapped in a composite double well (CBJJ) given by 
\begin{equation}\label{eq-dyn}
	V_{Trap}(x)=V_T(x)+V_0 \exp[-a(x+2)^2]+V_0 \exp[-a(x-2)^2]
\end{equation}
where
\begin{equation}\label{cdw}
	V_T(x) = \left\{
	\begin{matrix}
		\frac{1}{2}(x + 2)^2 , \hspace*{1cm} x < -\frac{1}{2} \cr 
		\frac{3}{2}(1-x^2) , \hspace*{1cm} |x| \le \frac{1}{2} \cr 
		\frac{1}{2}(x - 2)^2 , \hspace*{1cm} x > \frac{1}{2} \cr  
	\end{matrix}
	\right.\,.
\end{equation}  
Note that the trapping potential Eq.~\ref{eq-dyn} reduces to a symmetric double well (BJJ) for $V_0 =0$ and gives localization in the four-well trap for very large $V_0$. In this work, we will vary $V_0$ from $V_0=0$ to a sufficiently large value $V_0=10$ enough to reveal the growth of many-body signatures in the dynamics. The typical shape of the composite double well is shown in Fig.~\ref{fig-trap}. Initially, the system is prepared in the ground state of
\begin{equation}\label{eq-ini}
V_{Trap}(x)=\frac{1}{2}(x + 2)^2 +V_0 \exp[-a(x+2)^2]
\end{equation}
and at $t=0$, the system is quenched by suddenly changing the trap $V_{Trap}(x)$ from Eq.~(\ref{eq-ini}) to Eq.~(\ref{eq-dyn}).  We then simulate the post-quench out-of-equilibrium dynamics by solving the time-dependent many-body Schr\"odinger equation using MCTDHB method~\cite{Streltsov2007, Ofir2008}:

\begin{gather}\label{MBSE}
\hat H \Psi = i \frac{\partial \Psi}{\partial t},\\ \nonumber
\hat H(x_1,x_2,\ldots,x_N) =  \sum_{j=1}^{N} \hat h(x_j) +  \sum_{k>j=1}^N \hat{W}(x_j-x_k).    
\end{gather}

Here $x_j$ {is} the coordinate of the $j$-th boson, $\hat h(x) = \hat T(x) + \hat{V}_{Trap}(x)$ is the one-body Hamiltonian containing kinetic energy $T(x)$ and a trapping potential $V_{Trap}(x)$ terms, {and}
\textcolor{black}{the pairwise interaction between the $j$-th and $k$-th bosons is given by $W(x_j-x_k)=\lambda_0 \delta(x_j - x_i)$, $\lambda_0$ being the interaction strength}. Dimensionless units are employed throughout this work \textcolor{black}{by scaling the Hamiltonian by $\frac{\hbar^2}{mL^2}$ where $L$ is the length of the CBJJ  and $m$ is the mass of the boson. We adopt natural units, where $\hbar=m=1$.}

MCTDHB has already been extensively used in the literature 
\cite{Sakmann2009,MCTDHB_OCT,MCTDHB_Shapiro,Tunneling_Rapha,Axel2016,Axel2017,NJP2018Axel,NJP2019Axel,Cosme2017,NJP2015Schmelcher,NJP2017Camille,NJP2017Schmelcher, Haldar2018,  Haldar2019, ofir_review2020, Bhowmik2020, Bhowmik2022arxiv, Dutta2023}. A detailed discussion on MCTDHB can be found in~\cite{Ofir2008, ofir_review2020}. \textcolor{black}{In this method, the ansatz is taken as the superposition of all possible
	$\begin{pmatrix} 
	N+M-1\\
	N
	\end{pmatrix}$
configurations, obtained by distributing $N$ bosons in $M$ time-dependent single{-}particle orbitals $\phi_k(x,t)$, i.e,}
\begin{equation}
\label{MCTDHB_Psi}
\left|\Psi(t)\right> = 
\sum_{\vec{n}}C_{\vec{n}}(t)\left|\vec{n};t\right>,    
\end{equation}
where the occupations $\vec{n}=(n_1,n_2,\cdots,n_M)$ preserve the total number of bosons $N$. Although $M$ should be infinitely large for an exact calculation, one needs to truncate the series Eq~(\ref{MCTDHB_Psi}) at a finite $M$ in all practical numerical computations. In actual calculations, we keep on increasing $M$ until we reach the convergence for $M$, thereby obtaining a numerically exact result. \textcolor{black}{Here we would like to point out that the flexibility of incorporating as many $M$ as required is an advantage of MCTDHB over other popular many-body methods such as Bose-Hubbard model. Accordingly, we truncate Eq~(\ref{MCTDHB_Psi}) at $M=4$ for our present work on CBJJ [see Appendix for the convergence with respect to $M$] whereas $M=2$ is enough for BJJ~\cite{Haldar2018, Haldar2019}. This may give the impression that the number of orbitals $M$ required for numerically exact results is equal to the number of wells present in the trapping potential. However, there is no such direct correlation between the two. The number of orbitals $M$ required for achieving numerical convergence and hence a numerically exact description of the system depends on the degree of fragmentation of the condensate. In the quench dynamics, the system develops fragmentation with time depending on various factors such as the nature of the interaction, its strength, and trap geometry. However, for a weakly interacting system, it is routinely observed that the number of orbitals $M$ required for convergence is equal to the number of wells in the trapping potential.}  

Without going into further details, here we just mention the working equations of MCTDHB for determining the time-dependent coefficients $\{C_{\vec{n}}(t)\}$ 
and the time-dependent orbitals $\{\phi_k(x,t)\}$  
\begin{eqnarray}\label{MCTDHB1_equ}
&&i\left|\dot\phi_j\right\rangle  =  \hat {\mathbf P} \left[\hat h \left|\phi_j\right\rangle  + \sum^M_{k,s,q,l=1} 
  \left\{\rho(t)\right\}^{-1}_{jk} \rho_{ksql} \hat{W}_{sl} \left|\phi_q\right\rangle \right]; \nonumber\\
&&\hat {\mathbf P} = 1-\sum_{j^{\prime}=1}^{M}\left|\phi_{j^{\prime}}\left>\right<\phi_{j^{\prime}}\right| \nonumber\\
&&{\mathbf H}(t)C(t) = i\frac{\partial C(t)}{\partial t}.
\end{eqnarray}
where $\rho(t)$ is the reduced one-body density matrix [see Eqn.~(\ref{1RDM}) below], $\rho_{ksql}$ are the elements of the two-body reduced density matrix [see Eqn.~(\ref{2RDM})]~\cite{Lowdin, RDMs}, and ${\mathbf H}(t)$ is the Hamiltonian matrix $H_{\vec{n}\vec{n}'}(t) = \left<\vec{n};t\left|\hat H\right|\vec{n}';t\right>$. \textcolor{black}{Note that here, the working equations Eq.~(\ref{MCTDHB1_equ}) are expressed in terms of the reduced one-body and two-body reduced density matrices for having a compact look. Given the normalized many-body wavefunction $\Psi(t)$, the reduced one-body density matrix is given as
\begin{eqnarray}
\label{1RDM}
\rho^{(1)}(x_1|x_1^{\prime};t) & = &
N \int d x_2 \ldots d x_N \, \Psi^\ast(x_1^{\prime},x_2,\ldots,x_N;t)  \nonumber \\
& \, & \times \Psi(x_1,x_2,\ldots,x_N;t) \nonumber \\
& = &\sum_{j=1}^{M} n_j(t) \, \phi^{\ast{NO}}_j(x_1^{\prime},t)\phi^{NO}_j(x_1,t).
\end{eqnarray}
Here, $\phi^{NO}_j(x_1,t)$ are the time-dependent natural orbitals and $n_j(t)$ the time-dependent natural occupation numbers. 
The natural occupations $n_j(t)$ are used to characterize the {(time-varying)} degree of condensation in a system of interacting bosons \cite{PeO56} and satisfy $\sum_{j=1}^{M} n_j = N$. 
If only one {macroscopic} eigenvalue 
$n_1(t) \approx {\mathcal O}(N)$ exists, the system is condensed \cite{PeO56} whereas if 
there are more than one {macroscopic} eigenvalues, the BEC
is said to be fragmented {\cite{NoS82, No96, Spekkens99,Ueda}}. The diagonal of the $\rho^{(1)}(x_1|x_1^{\prime};t)$ gives the density of the system
$\rho(x;t) \equiv \rho^{(1)}(x|x^{\prime}=x;t) \label{1RDMdiag}$}.

\textcolor{black}{Similarly, the two-body density can be expressed as  
\begin{equation}\label{2RDM}
\begin{split}
\rho^{(2)}(x_{1},x_{2} \vert x_{1}^{\prime}, x_{2}^{\prime};t)= \hspace*{4cm}\\
N(N-1)\int_{}^{}d x_{3} \ldots d x_{N} \Psi^{*}(x_{1}^{\prime},x_{2}^{\prime},x_{3},\ldots,x_{N};t) \\
\times \Psi(x_{1},x_{2},x_{3}, \ldots, x_{N};t).
\end{split}
\end{equation} 
Therefore, the matrix elements of the two-body reduced density matrix are given by 
$\rho_{ksql}=\left<\Psi\left|b_k^\dag b_s^\dag b_q b_l\right|\Psi\right>$ where $b_k$ and $b_k^\dag$
are the bosonic annihilation and creation operators, respectively.}

\section{Results \& discusson}~\label{result}
In this section, we present our findings of the study on the many-body dynamics of an interacting bosonic gas in a composite double well as defined in Eq.~(\ref{eq-dyn}). Our main purpose of this study is to explore the growing role of higher energy bands as the system's complexity increases. Accordingly, as already mentioned above, we would vary $V_0$ from $V_0=0$ (for which the system is two-fold fragmented for $\Lambda = 0.1$ considered here) to $V_0=10$ when the system becomes four-fold fragmented (see below) and highlight 
the characteristics exclusively due to the participation of the higher energy bands. Additionally, we provided strong numerical evidence that the universality of degrees of fragmentation, earlier reported for two-fold fragmentation~\cite{Haldar2018, Haldar2019}, still takes place for more than $M=2$ time-dependent modes.

\subsection{Many-body dynamics}
As mentioned earlier, initially we prepared the system in the ground state of the left well [Eq.~(\ref{eq-ini})] of the CBJJ. With increasing values of $V_0$,  the ground states acquire increasingly complicated shapes which makes the dynamics further complex. To highlight this increasingly complex nature of the dynamics and the necessity of considering higher orbitals beyond $M=2$ to capture the same faithfully, here we present our findings of the temporal evolution of the survival probability, occupation numbers, and the many-particle position and momentum variances for various $V_0$ keeping $a=10$ fixed. \textcolor{black}{Note that $V_0$ mainly controls the height of the barrier in each well whereas $a$ primarily controls the intra-well barrier width, see Fig.~\ref{fig-trap}. Therefore, the impact of $V_0$ on the fragmentation should be more prominent than $a$. Accordingly, we have kept $a$ fixed to some intermediate value $a=10$ where the intra-well barrier is neither very narrow nor very wide. The qualitative physics described in this work will not change with $a$.}

\textcolor{black}{Furthermore, it is desired to use the same reference time scale while comparing the many-body tunneling dynamics in the composite double well with various barrier heights $V_0$.  While Rabi period $t_{Rabi}$ provides the natural time scale of the system, it varies with the trap geometry. However, $t_{Rabi}$ still would be of the same order of magnitude for all cases. The Rabi period for the BJJ, $t_{Rabi} = 132.498$ ~\cite{Haldar2018} suggests that the time period of the inter-band oscillations in the CBJJ should be of the order of a few $100$. Accordingly, we have scaled the time $t$ by $t_0=100$ while comparing the dynamics for various trap geometry corresponding to different $V_0$. This time $t$ in the dimensionless unit can be converted into standard time unit (such as second) by multiplying $t$ with the inverse of the scaling factor $mL^2/\hbar$ [see Sec.~\ref{method}]. Considering $L=1\mu$m and mass of the $^{87}$Rb $m=1.4431\times 10^{-25}$ kg, this conversion factor turns out to be $mL^2/\hbar=1.37$ milli second.}

\subsubsection{Survival probability}\label{pl}
\begin{figure}[!h]
\begin{center}
\subfloat[]{\includegraphics[width=0.45\textwidth]{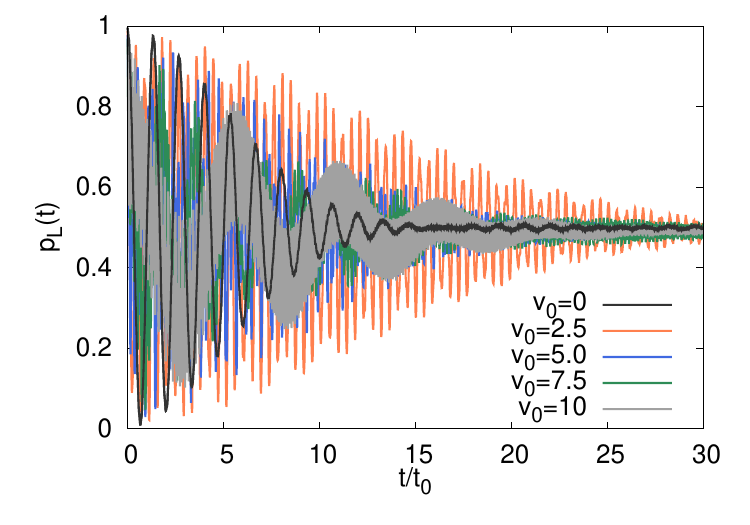}}
\subfloat[] {\includegraphics[width=0.45\textwidth]{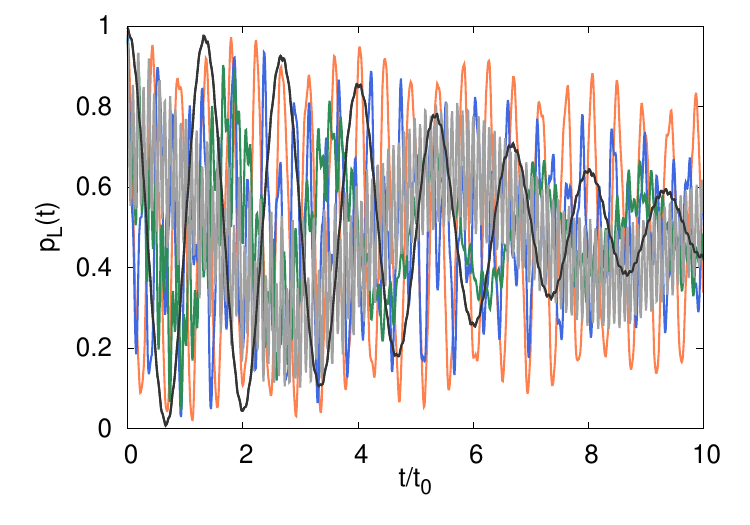}}
\vglue -0.5truecm
\end{center} 
\caption{Plot of the $p_L(t)$ as a function of time $t$ for various barrier heights $V_0$ [panel (a)]. The magnified view of the first few oscillations of $p_L(t)$ is shown in panel (b). Colour codes are explained in panel (a). These results are obtained with $M=4$ orbitals for $N=100$ particles with interaction strength $\Lambda=0.1$. }
\label{fig-pl}
\end{figure}

The survival probability $p_L(t)$, which is a measure of density oscillations between the composite double wells, can be defined as
\begin{equation}
	p_L(t)=\int_{-\infty}^0 {\rm d}x \frac{\rho(x;t)}{N},
\end{equation}
where $\rho(x;t)$ is the density in the left composite well. It is closely connected to the population imbalance which is routinely studied in the experiments.  In Fig.~\ref{fig-pl}, we plot the survival probability $p_L(t)$ in the left well for different values of $V_0$. For $V_0=0$,  we get back the standard BJJ comprising of two symmetrical wells connected through a barrier, and accordingly, $p_L(t)$ also exhibits the usual collapse of density oscillations with time. As $V_0$ is increased from zero, a hump appears in either well of the BJJ and it turns into the composite BJJ. We observe that irregularity appears in the oscillations of the $p_L(t)$ for $V_0 \neq 0$. Further, the irregularities enhance with the increase in $V_0$. With further increase in $V_0$, we observe two distinct oscillations for $V_0=10$. The high-frequency oscillations indicate the tunneling between the two adjacent shallow wells. In contrast, the low-frequency oscillations, appearing as the envelope to the high-frequency oscillations, can be attributed to the tunneling of the whole system between the two composite double wells. The emergence of two distinct frequencies in the tunneling dynamics implies atomic transitions between two distinct sets of energy levels which means more than one energy band is involved. \textcolor{black}{While the fast oscillations of short time period correspond to inter-band atomic transitions, the slower oscillations with longer time period are due to the intra-band atomic transitions. To corroborate this, ideally one should calculate the many-body energy levels of the CBJJ. However, calculating even the low-lying many-body excited states of such systems is extremely challenging. While efforts to develop a linear response theory to MCTDHB, called LR-MCTDHB~\cite{LR-MCTDHB2012}, are on to calculate the low-lying excitations, such computations are quite time-consuming, and achieving numerical convergence can be too challenging for complex systems such as the CBJJ. However, at least for weak interactions, the appearance of two distinct oscillations can be understood from the non-interacting picture. At the lower part of the spectrum of a composite double well, the closely lying energy levels form a band while in the upper part of the spectrum, the effect of the barriers is less pronounced and the spectrum is similar to that of a wide harmonic well. Therefore there will be two different types of oscillations arising from atomic transitions between the energy levels of an energy band and between energy levels belonging to two different energy bands. In particular, for $V_{0}=10$, the intra-band time period for transitions between the energy levels of the lowest energy band is about $T_{intra-band} \sim 500$  while the intra-band transitions have a time period $T_{inter-band}$ of the order 10 for the transitions between the lowest band and the first higher band. This sharp difference in the time scale of two oscillations is clearly visible in Fig~\ref{fig-pl}. As  $V_0$ is reduced,  the gaps between the energy bands reduce which decreases the difference in the time scale of $T_{inter-band}$ and $T_{intra-band}$.} Therefore, two oscillations in Fig~\ref{fig-pl} are not distinctly noticeable for smaller barrier heights $V_0$.  
The decay in the density oscillations is a purely many-body phenomenon and cannot be captured by mean-field theory~\cite{Haldar2019}. The different decay rates of the density oscillations depending on the values of $V_0$ come from the initial shape of the ground orbital.

\subsubsection{Occupation numbers}
Fragmentation of BEC provides information about the relative macroscopic populations in higher orbitals. The connection between the damping of density oscillations and the fragmentation for a BJJ is already well demonstrated in Ref.~\cite{Sakmann2009}.  Naturally, our findings for the $p_L(t)$ mentioned above demand a study of the fragmentation as a function of time for various heights of the hump $V_0$. This would help us to correlate the changes in density oscillations with the participation of higher energy bands. 

The development of fragmentation in the system is characterized by the time evolution of the natural occupations per particle $\frac{n_i}{N}$.  The system is said to be condensed if $\frac{n_1}{N} \sim \mathcal{O}(1)$.  On the other hand, the system is fragmented if $\frac{n_i}{N} \sim \mathcal{O}(1)$ for more than one orbital.

\begin{figure}[!h]
\begin{center}		
\includegraphics[width=0.9\textwidth]{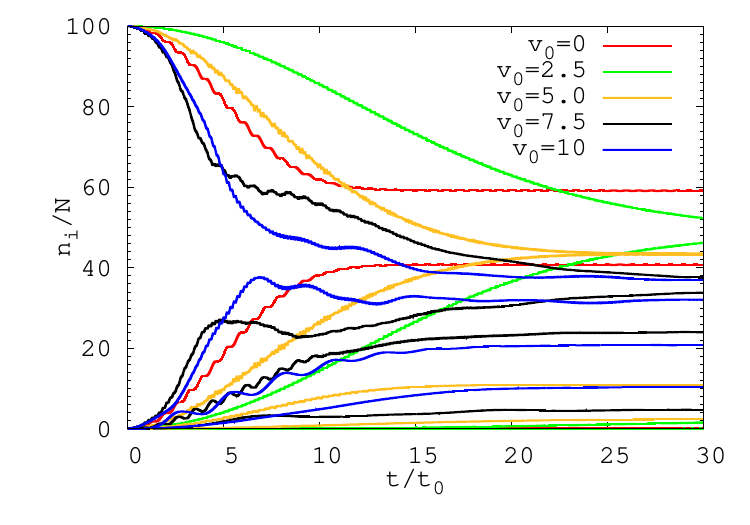}
\end{center} 
\vglue -0.50truecm
\caption{Plot of $\frac{n_i}{N}$ $(i=1,2,3,$ and   $4)$ for standard BJJ $(V_0=0)$ and the composite BJJ with various barrier heights $V_0$. These results are obtained with $M=4$ orbitals for a system of $N=100$ particles with interaction strength $\Lambda=0.1$.}
\label{fig-no}
\end{figure}

We present our results in Fig.~\ref{fig-no}. It is found that at $t=0$, for all the values of $V_0$, only the ground orbital is occupied. It signifies that although the shape of the ground state is modified due to different values of $V_0$, the coherency of it is unchanged. At $t>0$, we find that the system gradually becomes fragmented for all $V_0$. While for BJJ ($V_0=0$) the system becomes two-fold fragmented, the degree of fragmentation increases for $V_0 \neq 0$.  For BJJ ($V_0 = 0$), after a sufficiently long time, the system becomes two-fold fragmented with $\frac{n_1}{N} \approx 60\%$ and $\frac{n_2}{N} \approx 40\%$ while the occupations in all higher orbitals are negligibly small. Therefore, a BJJ, with   interaction strength cosidered,  can be accurately described with only $M=2$ orbitals. However, the occupations in the third and fourth orbitals start increasing with an increase in $V_0$ at the cost of occupations in the first and second orbitals. While for  $V_0=2.5$, we see $\frac{n_3}{N} \approx 5\%$, it grows to over $\frac{n_3}{N} \approx 20\%$ for $V_0=10$. Similarly, $\frac{n_4}{N}$ which is very small for $V_0=2.5$ grows to $\frac{n_4}{N} \approx 5\%$ for $V_0=10$. Accordingly, the occupations in the first orbital reduces from $\frac{n_1}{N}\sim 60\%$ for $V_0=0$ to $\frac{n_1}{N}\sim 50\%$ for $V_0=2.5$, $\frac{n_1}{N}\sim 45\%$ for $V_0=5$, $\frac{n_1}{N}\sim 37\%$ for $V_0=7.5$. With further increase in $V_0$ to $V_0=10$, there is no significant variation in $\frac{n_1}{N}$. Similarly, $\frac{n_2}{N}$ first increases from $\frac{n_2}{N} \sim 40\%$ for $V_0=0$ to $\frac{n_2}{N} \sim 50\% $ for $V_0=2.5$, and then gradually decreases to $\frac{n_2}{N} \sim 45\% $ for $V_0=5$, $\frac{n_2}{N} \sim 35\% $ for $V_0=7.5$, and finally to $\frac{n_2}{N} \sim 32\% $ for $V_0=10$.  So, initially, primarily the second orbital gains population at the cost of the first orbital, and both orbitals achieve similar occupations for $V_0=2.5$. With further increase in  $V_0$, particles are transferred to the third and fourth orbitals from the first two orbitals. So, for $V_0=5$ the occupations in the first and second orbitals become closer and at the same time occupations in the third and fourth orbitals also increase. While there is a further reduction in occupations in both of the first two orbitals, more particles transfer from the second orbital resulting in reduced occupations in the second orbital. Finally, as $V_0$ increases to $V_0=10$ from $V_0=7.5$, there is further redistribution of particles among the second, third, and fourth orbitals so that  $\frac{n_1}{N}$ remains practically the same.  Therefore it is evident that there is a redistribution of the particles among the orbitals and finally, for significantly large $V_0$ there is a more equitable distribution of the population. This indicates the greater role played by the higher orbitals in the tunneling dynamics of the CBJJ ($V_0 \neq 0$).  

In this connection, it is instructive to note that the density oscillations die out in BJJ as the system becomes correlated. However, the damping slows down for the CBJJ even though the CBJJ is more fragmented than BJJ. This may be attributed to the delay in the growth of fragmentation in CBJJ for smaller values of $V_0$. But for $V_0 \ge 7.5$, the damping is slower even though the correlation in the system grows faster than BJJ. This shows the complicated relationship between the damping of density oscillations and the growth of correlation in the system in the case of CBJJ.   

\subsubsection{Many-particle variances}
Next, we consider the many-particle position and momentum variances of the system, which is a measure of the quantum resolution of measurement of any observable. Although these cannot be measured easily, these are fundamental quantities due to their connection with the uncertainty principle. Contrary to fragmentation, which reflects only relative occupations, many-body variances are dependent on the actual number of fragmented atoms. Therefore, these are expected to bear more prominent signatures of the actual occupations in higher orbitals and thereby may throw more light on the role of higher orbitals in the dynamics. 

\begin{figure}[!h]
\centering
  \subfloat[] { \includegraphics[width=0.45\textwidth]{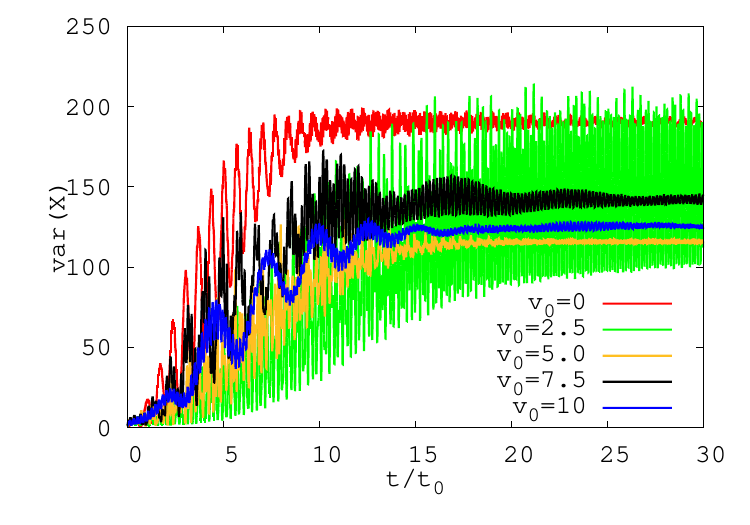}}
   \subfloat[]  {\includegraphics[width=0.45\textwidth]{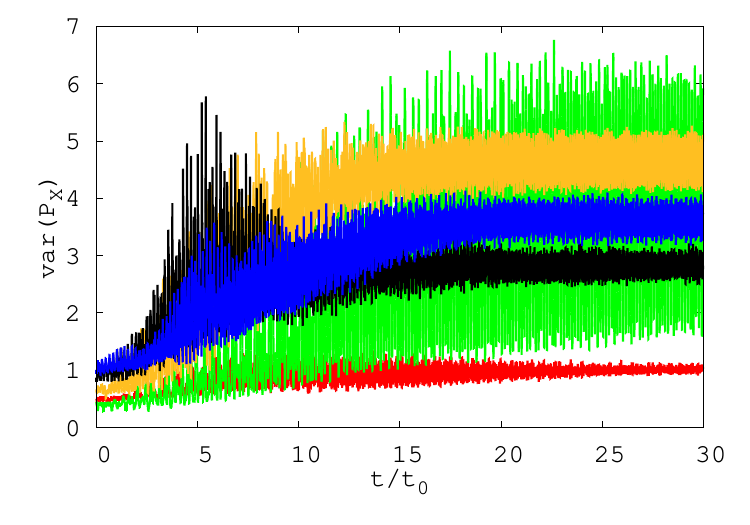}} 
\caption{Plot of the many-particle variances of position [panel (a)] and momentum [panel (b)] operators for various barrier heights $V_0$. Here the results are obtained with $M=4$ orbitals for $N=100$ particles and interaction strength $\Lambda=0.1$.}
	\label{fig-var}
\end{figure}

One can calculate the variance per particle $\frac{1}{N}\Delta_{\hat A}^2(t)$ for any many-body operator $\hat A=\sum_{j=1}^N \hat a(x_j)$, which is obtained from single-particle Hermitian operator $\hat a(x_j)$, by 

\begin{gather}\label{dis}
\frac{1}{N}\Delta_{\hat A}^2(t)   =  \frac{1}{N} 
\left[\langle\Psi(t)|\hat A^2|\Psi(t)\rangle - \langle\Psi(t)|\hat A|\Psi(t)\rangle^2\right]  \\ \nonumber
\equiv \Delta_{\hat a, density}^2(t) + \Delta_{\hat a, MB}^2(t),  \\ \nonumber
\Delta_{\hat a, density}^2(t) = 
\int d x \frac{\rho(x;t)}{N} a^2(x) - \left[\int dx \frac{\rho(x;t)}{N} a(x) \right]^2, \\   \nonumber
 \Delta_{\hat a, MB}^2(t)  = \frac{\rho_{1111}(t)}{N} \left[\int d x |\phi^{NO}_1(x;t)|^2 a(x) \right]^2\\ \nonumber
 - (N-1) \left[\int dx \frac{\rho(x;t)}{N} a(x) \right]^2\\ \nonumber
 + \sum_{jpkq\ne 1111}\frac{\rho_{jpkq}(t)}{N} \left[\int d x \phi^{\ast{NO}}_j(x;t) \phi^{NO}_k(x;t) a(x) \right] \\ \nonumber
 \times \left[\int dx \phi^{\ast{NO}}_p(x;t) \phi^{NO}_q(x;t) a(x)\right].
\end{gather} 
Accordingly in Fig.~\ref{fig-var} (a), we present our results for the many-particle position variance while  Fig.~\ref{fig-var} (b) depicts the results for the many-particle momentum variance.

For $V_0=0$, we again get back the oscillatory growth of  $\frac{1}{N}\Delta_{\hat X}^2$ with time followed by the saturation as the density oscillation collapses. For $V_0 \neq 0$, we get the overall behaviour quite similar to the BJJ  but the main difference lies in the details. For a small value of $V_0=2.5$, the growth rate of  $\frac{1}{N}\Delta_{\hat X}^2$ becomes slower which is consistent with the fragmentation dynamics. However, the fluctuations in the saturation value are very high for such a small value of $V_0$. \textcolor{black}{This is expected because all three barriers are of similar heights [see Fig.~\ref{fig-trap}] and therefore, the system is delocalized over all four wells. The fluctuations are expected to be maximum when all three barriers are of the same height which should occur for $V_0 = 1.5$.}  With a further increase of $V_0$, the growth rate of $\frac{1}{N}\Delta_{\hat X}^2$ increases but still remains slower than BJJ. Also, the fluctuations reduce drastically for higher values of $V_0$.

The fluctuations in $\frac{1}{N}\Delta_{\hat X}^2$ about the saturation value are the manifestation of the breathing oscillations within the two wells of each of the compound wells. For small intra-well barrier height $V_0$, large breathing oscillations between the two wells of each compound wells lead to large fluctuations in $\frac{1}{N}\Delta_{\hat X}^2$. With increasing $V_0$, the breathing oscillations get damped leading to smaller fluctuations in $\frac{1}{N}\Delta_{\hat X}^2$ for higher values of $V_0$.

The dynamics of atoms will have a more prominent impact on the many-particle momentum variance. Naturally, we again observe fluctuations in $\frac{1}{N}\Delta^2_{\hat{P}_X}$ around a mean value for BJJ ($V_0=0$). For a small $V_0$, \textcolor{black}{the barrier heights of both the inter-well and intra-well barriers are again of similar heights resulting in large intra-well breathing oscillations in addition to inter-well tunneling oscillations. Accordingly, we observe very large fluctuations in $\frac{1}{N}\Delta^2_{\hat{P}_X}$.} With further increase in $V_0$, the damping in the intra-well breathing oscillations results in the gradual reduction in fluctuations of $\frac{1}{N}\Delta^2_{\hat{P}_X}$. For such values of $V_0$, $\frac{1}{N}\Delta^2_{\hat{P}_X}$ fluctuates around a mean value which initially increases slightly with time before stabilising to a saturation value. This saturation value also depends on $V_0$ and first increases before gradually decreasing with increasing $V_0$. 

\subsection{Universality of fragmentation}\label{univ}
\begin{figure*}[!h]
\begin{center}
 \subfloat[]{\includegraphics[width=0.45\textwidth]{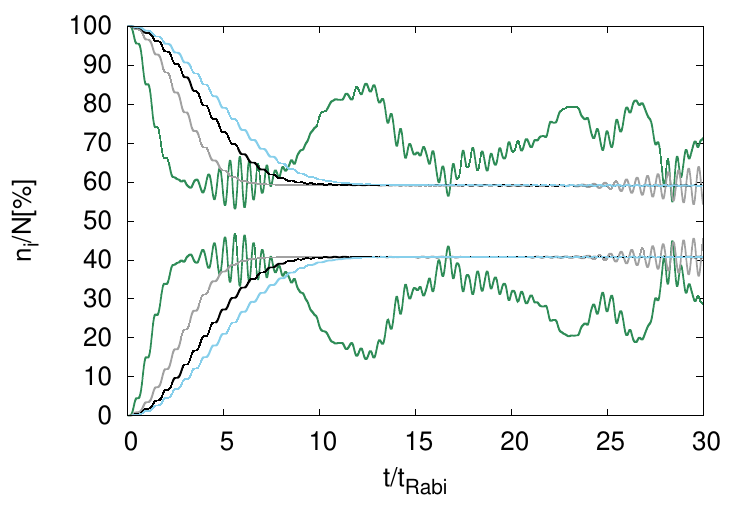}}
  \subfloat[]{\includegraphics[width=0.45\textwidth]{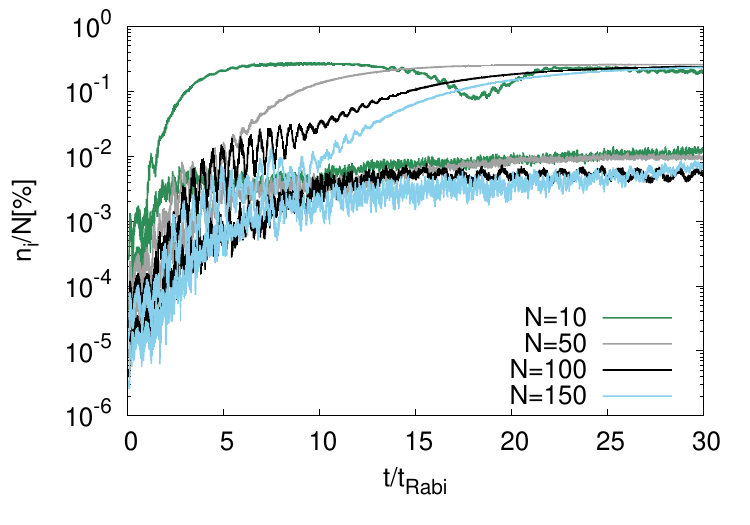}}\\
  \subfloat[]{\includegraphics[width=0.45\textwidth]{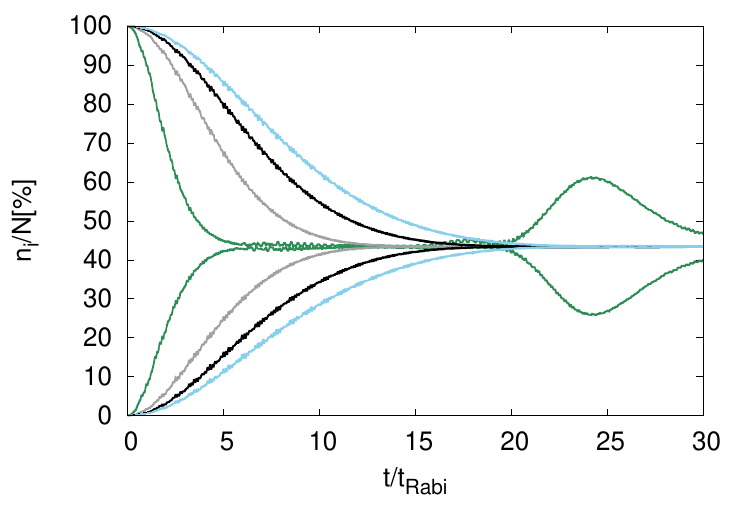}}
\subfloat[]{\includegraphics[width=0.45\textwidth]{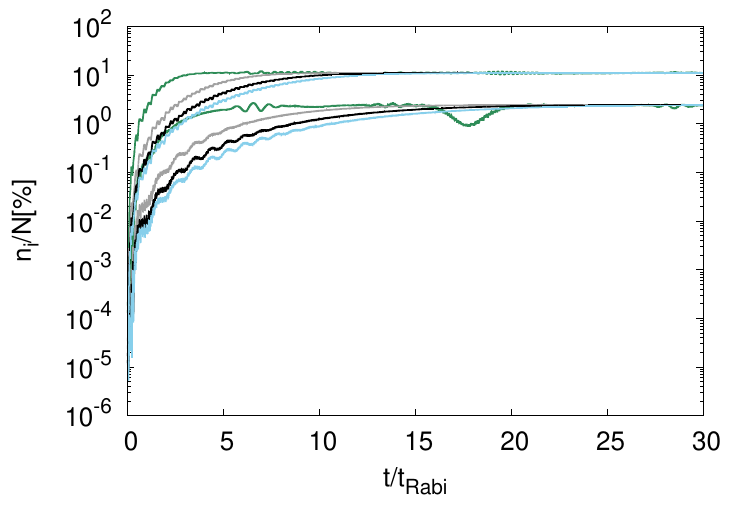}}\\
   \subfloat[]{\includegraphics[width=0.45\textwidth]{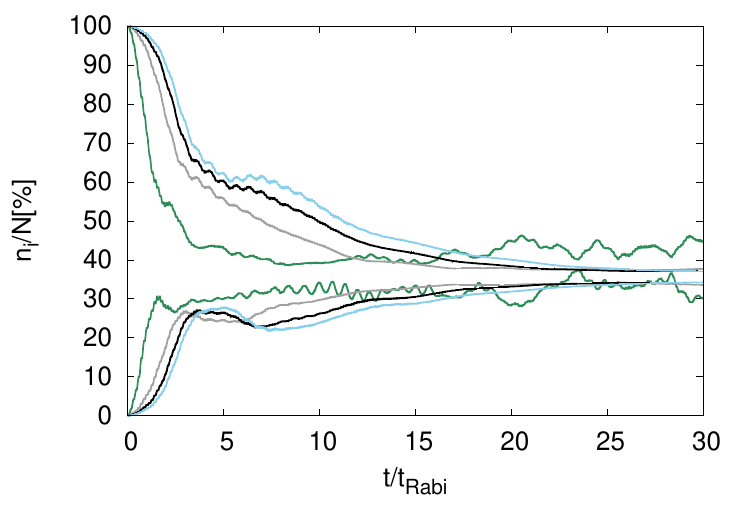}}
   \subfloat[]{\includegraphics[width=0.45\textwidth]{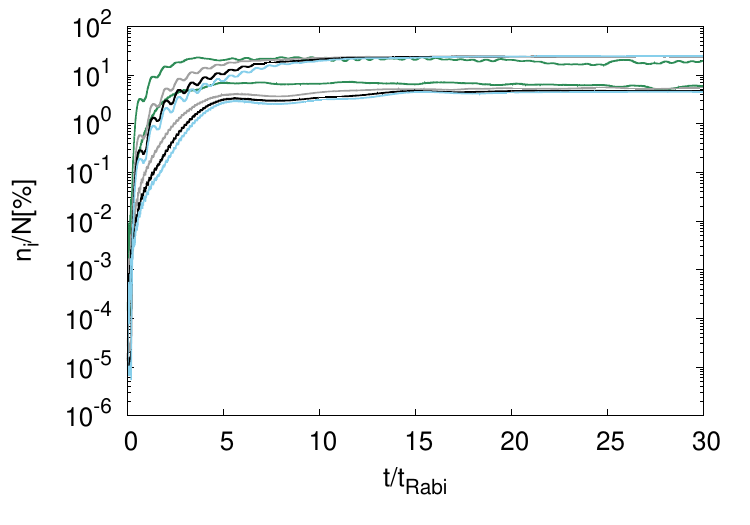}}\\
\end{center}
\caption{Universality of fragmentation in a standard BJJ (i.e. $V_0=0$) [panel (a) \& panel (b)] and composite BJJ with $V_0=5$ [panel (c) \& panel (d)], and $V_0=7.5$ [panel (e) \& panel (f)]. In each row, the left panels show the occupations for the first two orbitals while the right panels show the occupation for the third and fourth orbitals. Since the occupations in higher orbitals are small, these are shown in the log scale. Colour codes are explained in panel (b). Results shown here are obtained with $M=4$ for interaction strength $\Lambda = 0.1 $.}
\label{fig-uni}
\end{figure*}

Since the participation of the higher energy bands significantly affected the growth of fragmentation in the system, it is natural to ask about its impact on the universality of the degree of fragmentation with respect to $N$. It is a novel phenomenon predicted for the BJJ dynamics when only a single band plays the dominant role in the dynamics. Therefore, next, we examine the degree of fragmentation of the CBJJ when the second band (in addition to the lowest band) also plays a significant role in the dynamics. We present our results for different $V_0$  in Fig.~\ref{fig-uni}. The panels in the left column show $\frac{n_i}{N}$ for the first two orbitals while those on the right column exhibit the same for the third and fourth orbitals. 

For the symmetric double well ($V_0=0$), we get back the usual two-fold fragmentation with the $\frac{n_1}{N} \sim 60\%$ and $\frac{n_2}{N} \sim 40\%$ for all $N$ keeping $\Lambda$ fixed at $0.1$. Although there are no macroscopic occupations in the higher orbitals for BJJ, we still observe the depletion of the same order in the third ($\frac{n_3}{N} \sim 10^{-3}$) and fourth ($\frac{n_4}{N} \sim 10^{-4}$) orbitals for all $N$.

For a finite $V_0$, the system is now four-fold fragmented. However, we still find that over time $\frac{n_i}{N}$ for all four orbitals reaches a fixed value for various $N$  corresponding to the same $\Lambda$  for all $V_0$ studied here.  However, as discussed above, the final values of $\frac{n_i}{N}$ change with an increase in $V_0$, and the particles of the system become more evenly distributed among the four orbitals with reduced occupations in lower orbitals and increased occupations in higher orbitals. Therefore, even though the details of fragmentation vary for various trap geometry, the systems exhibit the universality of fragmentation for all four orbitals with respect to $N$ corresponding to a fixed $\Lambda=0.1$ irrespective of the value of $V_0$.  However, we have noticed that the pace of appearance of the universality of fragmentation depends on $V_0$. As we switch on the internal barrier $V_0$, for a small value of $V_0$ it takes longer to observe the universality of fragmentation than the symmetric double well $(V_0=0)$. Then, as $V_0$ is increased, the universality appears faster. Furthermore, for a fixed value of $V_0$, the system fragments later for larger $N$. This is expected as the many-body effects are reduced with increasing $N$ keeping $\Lambda$ fixed.

In our earlier study~\cite{Haldar2019}, we demonstrated that the universality of fragmentation is preserved in asymmetric BJJ. So, our present study in combination with the earlier findings indicates that the universality of fragmentation with respect to $N$ is neither limited to systems where only two orbitals play a dominant role nor is susceptible to the geometry of the double well potential. Thus it is indeed a robust many-body effect.

\section{Conclusions}~\label{conclusion}
In this work, we studied the many-body tunneling dynamics of an interacting Bose gas in a composite double well. The composite double well is formed by merging two deformed harmonic wells which have a hump at their center. The complex geometry of the system manifests in the rich many-body tunneling dynamics involving higher energy bands. We examined the complexity of dynamics by analysing the time evolution of survival probability, occupation numbers, and many-particle variances. All these quantities were previously used to analyze the many-body features of BJJ dynamics. 

We found irregularity in the density oscillations for very small hump and then, two distinct oscillations for bigger hump. While the fast high-frequency oscillation is due to the tunneling through the hump within the same well, the low-frequency tunneling is due to the tunneling between the two wells. The emergence of two frequencies in the tunneling dynamics indicates the transitions between two different sets of energy levels and thereby involvement of more than one energy band.  
 
Indeed, we observed (see appendix) that at least four orbitals are required for the accurate description of the many-body dynamics of the system contrary to the standard BJJ dynamics where two orbitals are sufficient. 

The correlation between the damping of the density oscillation and the growth of fragmentation is already well-established. We observed that the same correlations still exist even for composite BJJ. However, the degree of fragmentation, as well as the growth of the fragmentation, significantly differ in the two cases. While the BJJ becomes two-fold fragmented, CBJJ is four-fold fragmented. On the other hand, the growth of fragmentation becomes slower in CBJJ except for very high hump $V_0$.  Accordingly, damping of density oscillations becomes slower in CBJJ. Furthermore, the occupations of different orbitals $\frac{n_i}{N}$ clearly show the more prominent participation of the third and fourth orbitals in the case of CBJJ. All these observations are further strengthened by the study of $\frac{1}{N}\Delta^2_{\hat{X}}$ and $\frac{1}{N}\Delta^2_{\hat{P}_X}$ which exhibit higher fluctuations but slower growth.

Another interesting finding of this study is that the universality of the fragmentation with respect to the occupation in an orbital corresponding to a fixed interaction strength $\Lambda$ is not limited to systems where only two orbitals play dominant roles. Furthermore, not only universality is observed for orbitals with macroscopic occupations but even orbitals with microscopic occupations also have the same order of occupations with respect to different $N$ but with a fixed $\Lambda$. And these features are observed for all $V_0$ which reaffirms that the universality of fragmentation is a global many-body phenomenon. 

Thus we see that depending on the systems, the role of a different number of orbitals in the dynamics becomes dominant. Therefore, it is necessary to have a many-body method capable of taking into account all relevant orbitals for an accurate description of the complex many-body dynamics. The MCTDHB method is an important and promising many-body technique in this direction. The present work can inspire us to investigate the tunneling dynamics of even more intriguing ground states of bosonic and fermionic ensembles.

\section{Appendix}

\subsection{Numerical convergence of dynamical quantities}
In our calculation, we truncated Eq.~(\ref{MCTDHB_Psi}) at $M=4$ orbitals [see Ref.~\cite{Ofir2008, ofir_review2020} for details] which is just sufficient for a numerically accurate description of the many-body dynamics. Considering a system of $N=10$ particles, here we explicitly demonstrate that $M=4$ is enough to achieve a good degree of numerical convergence of our results with respect to $M$.  We point out that the accuracy of our actual results is better than the accuracy shown in this section, as the overall convergence of the quantities improves with increasing $N$ for a fixed $\Lambda$~\cite{Lieb_PRA, Lieb_PRL, Erdos_PRL, MATH_ERDOS_REF}.

Before we discuss the convergence of our results for the dynamics, it is imperative to ensure the convergence of the initial states used for the time propagation. The initial state for the tunneling dynamics of the CBJJ is the ground state of the standard BJJ. 
It has already been demonstrated in previous studies~\cite{Sakmann2009, Haldar2018}, that $M=2$ orbitals are enough for an accurate description of the BJJ dynamics even with a long-range interaction. Here we computed the initial states with $M=4$ orbitals as we would need at least $M=4$ orbitals to study the dynamics of CBJJ. Therefore, at the onset of our study, we can rest assured about the convergence of our initial state with respect to $M$.

To demonstrate the numerical convergence of the dynamics, it is required to show the convergence of the time variation of $p_L(t)$, $\frac{n_i}{N}$, $\frac{1}{N}\Delta_{\hat X}^2$ and $\frac{1}{N}\Delta_{\hat P_X}^2$. However, it is well demonstrated that achieving convergence for the many-particle variances of position and momentum operators is more challenging. Therefore, the demonstration of convergences for these quantities should automatically demonstrate the convergence of $p_L(t)$ as well as $\frac{n_i}{N}$. Still, we will explicitly show the convergence of occupation numbers in addition to the convergence of the many-particle variances, since the characteristics of the time evolution of the occupation numbers are directly related to the features of the many-particle variances of position and momentum operators. Moreover, demonstration of convergence for systems with a higher degree of fragmentation would automatically imply convergence for systems with a lower degree of fragmentation as the many-body effects are diminished with the decrease in the occupations in higher orbitals. Therefore, demonstrating the convergence of various quantities for $V_0 = 10$ would automatically imply the convergence for other $V_0$ considered in this work for which the system is less fragmented. We will discuss the convergence of these quantities with respect to $M$ for the duration of time studied in this work.

\subsection{Convergence of fragmentation and depletion}
\begin{figure}[!h]
\begin{center}		
\includegraphics[width=0.9\textwidth]{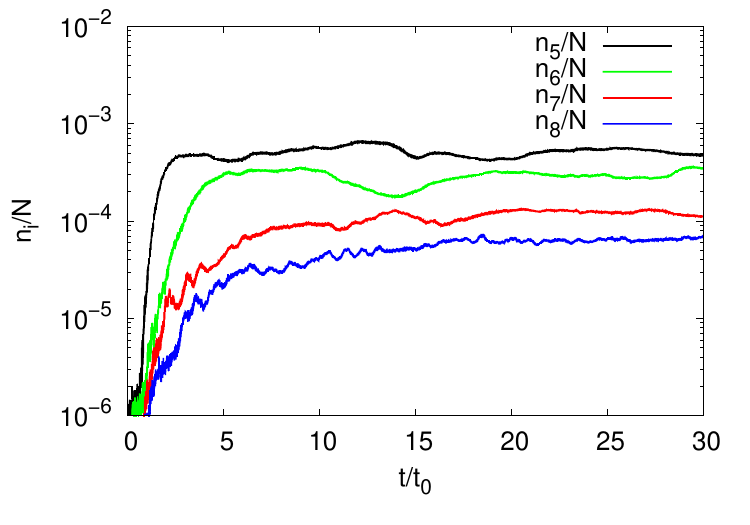}
\end{center} 
\vglue -1.0truecm
\caption{Plot of occupations in higher $M > 4$ orbitals viz. $\frac{n_i}{N}$ $(i=5,6,7, \mathrm{and} 8)$ for the composite BJJ with $V_0 = 10$.}
\label{fig-depl}
\end{figure}

We start with the demonstration of convergence of the occupation numbers of different orbitals for the composite BJJ. Although the demonstration of convergence for $V_0=10$ is sufficient, we still explicitly show the convergence for $V_0=7.5$ in addition to $V_0=10$ since in section~\ref{univ} we showed the universality of fragmentation for $V_0$ up to $V_0=7.5$.  
\begin{figure*}[!h]
\centering
 \subfloat[]{\includegraphics[width=0.45\textwidth]{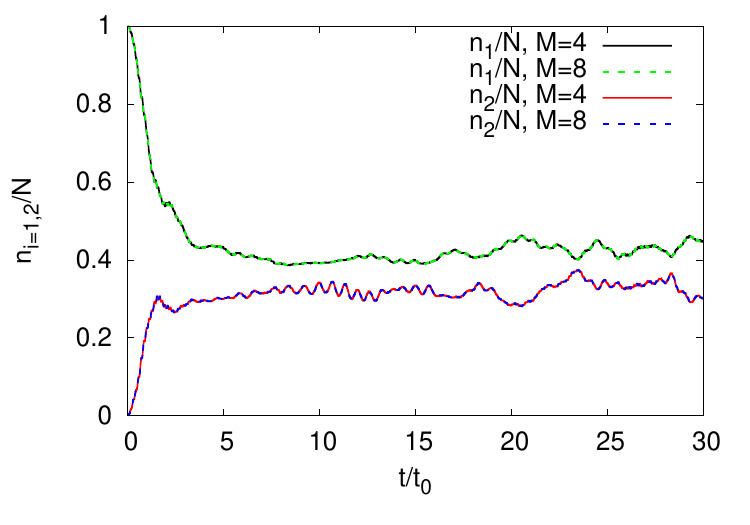}}
 \subfloat[]{\includegraphics[width=0.45\textwidth]{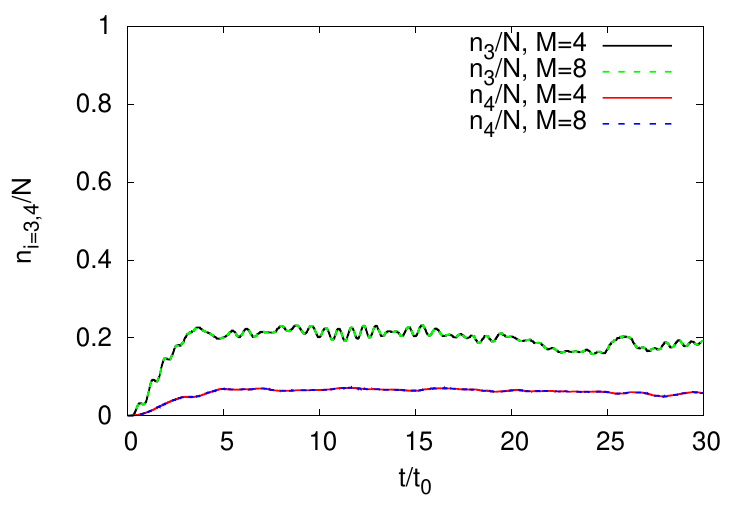}}\\  
 \subfloat[]{\includegraphics[width=0.45\textwidth]{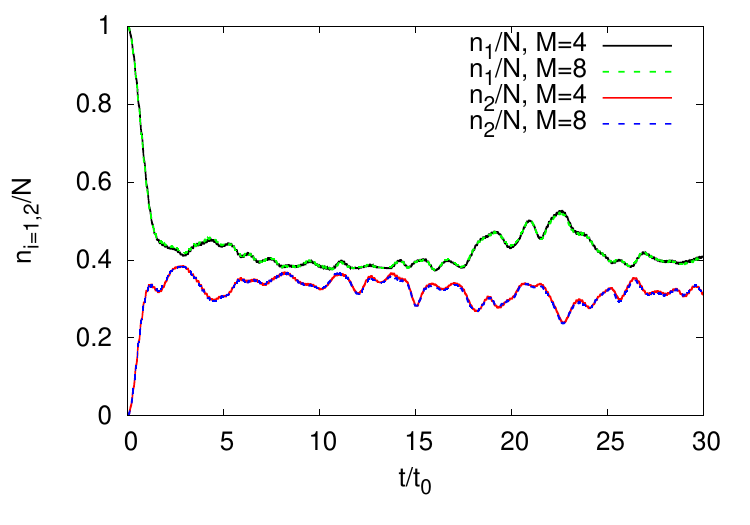}}
 \subfloat[]{\includegraphics[width=0.45\textwidth]{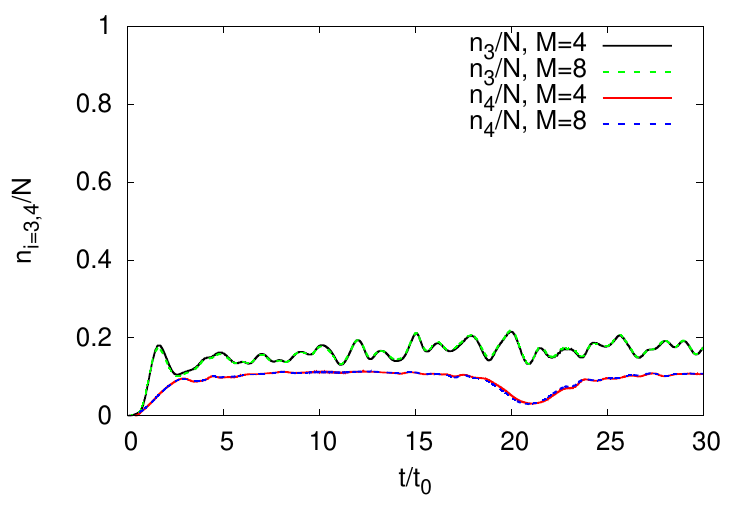}}
\caption{Plot of the convergence of the occupations of  (a) the first and second orbitals and (b) the third and fourth orbitals with respect to $M$ for a CBJJ of barrier height $V_0=7.5$, $N=10$, and  $\Lambda=0.1$. Convergence of occupation numbers for $V_0=10$, $N=10$, and  $\Lambda=0.1$ are shown in panel (c) (first and second orbitals) and (d) (third and fourth orbitals) respectively.}
\label{fig-frag-conv}
\end{figure*}\\
First, we consider the first four orbitals which have macroscopic occupations for CBJJ (though the third and fourth orbitals for BJJ still have microscopic occupations). In Fig.~\ref{fig-frag-conv} we plot $\frac{n_i}{N}[\%]$ for the first $4$ orbitals for different $M \ge 4$ to exhibit the convergence of our results. We observe that the results for  $M=4$ and $M=8$ essentially overlap with each other for both $V_0 = 7.5$ and $V_0 = 10$. Since the many-body results smoothly approach the mean-field limit as $N$ increases,  
our results for $N=100$ accurately demonstrate the development of fragmentation in the course of the out-of-equilibrium many-body dynamics.  Further, in Fig.~\ref{fig-depl}, we explicitly show that the occupations for the higher orbitals for $V_0 = 10$ are microscopically small.

\subsection{Convergence of many-particle position and momentum variances}
\begin{figure}[!h]
\centering  
 \subfloat[]{ \includegraphics[width=0.45\textwidth]{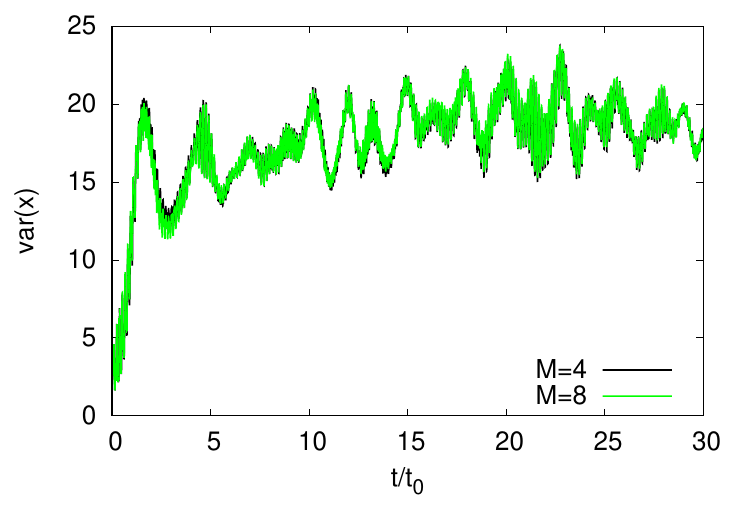}	}
 \subfloat[]{\includegraphics[width=0.45\textwidth]{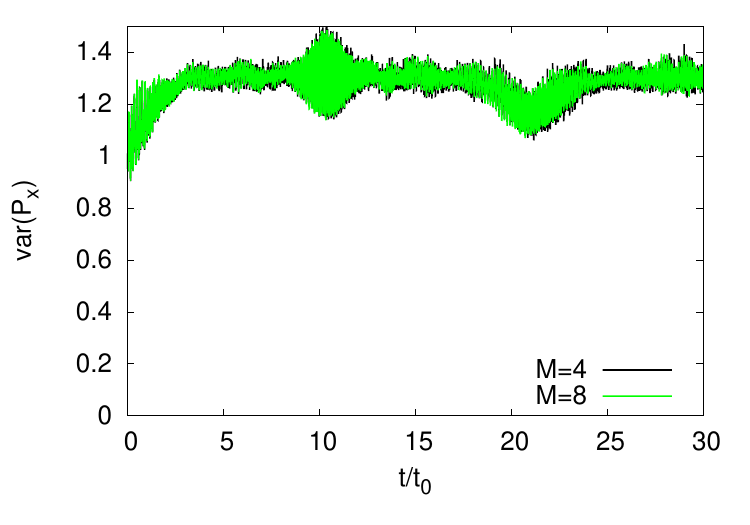}}
\caption{Plot of the convergence of (a) many-particle position variance and (b) many-particle momentum variance with respect to $M$ for a CBJJ of barrier height $V_0=10$, $N=10$, and  $\Lambda=0.1$.}
\label{fig-var-conv}
\end{figure}

Next, we present the convergence results for $\frac{1}{N}\Delta_{\hat X}^2$ and $\frac{1}{N}\Delta^2_{\hat{P}_X}$ [Fig~\ref{fig-var-conv}]. For both $M=4$ and $M=8$ orbitals, we observe the same qualitative feature that $\frac{1}{N}\Delta_{\hat X}^2$ increases with the growth of fragmentation and then exhibits an irregular oscillation [Fig~\ref{fig-var-conv}(a)]. In conformity with our findings of the fragmentation, here also we find a good degree of overlap between the fluctuations of $\frac{1}{N}\Delta_{\hat X}^2$ for $M=4$ and $M=8$. Similarly, we notice a very high degree of overlapping between $\frac{1}{N}\Delta^2_{\hat{P}_X}$ for $M=4$ and $M=8$ [Fig~\ref{fig-var-conv}(b)]. However, the deviation between the $\frac{1}{N}\Delta^2_{\hat{P}_X}$ for $M=4$ and $M=8$ is slightly higher than $\frac{1}{N}\Delta_{\hat X}^2$. This is expected as achieving convergence for $\frac{1}{N}\Delta^2_{\hat{P}_X}$ is more difficult. 

As the convergence significantly improves with increasing $N$, we can safely assume that the main findings for the system size $N=100$ accurately demonstrate the many-body features of the out-of-equilibrium dynamics of a fragmented system.  

\section*{Acknowledgments}
We acknowledge Prof. Ofir E Alon and Rhombik Roy for some valuable discussions. 
SKH acknowledges the support from the Department of Science and Technology (DST), India, through TARE Grant No.: TAR/2021/000136. Computation time on the High-Performance Computing system Hive of the Faculty of Natural Sciences at the University of Haifa and at the High-Performance Computing Center Stuttgart (HLRS) is gratefully acknowledged.

\reftitle{References}

\externalbibliography{yes}
\bibliography{ref}

\end{document}